\documentclass[epj]{svjour}

\usepackage{graphicx}
\usepackage{color}     
\usepackage{amsmath}   

\newcommand{\be}{\begin{equation}}
\newcommand{\ee}{\end{equation}}

\usepackage[normalem]{ulem}

\graphicspath{{figures/},.}
\begin{document}
\title{Role of Network Topology in the Synchronization of Power Systems}
\author{
Sergi Lozano \inst{1,2,3} 
\thanks{
\email{slozano@iphes.cat}
}
\and 
Lubos Buzna \inst{4}
\thanks{
\email{buzna@frdsa.uniza.sk}
}
\and
Albert D\'{\i}az-Guilera \inst{5}
\thanks{
\email{albert.diaz@ub.edu}
}
}                      
%
%
\institute{IPHES, Institut Catal\`a de Paleoecologia Humana i Evoluci\'o Social, 43003 Tarragona, Spain
\and
\`Area de Prehistoria, Universitat Rovira i Virgili (URV), 43002 Tarragona, Spain
\and
ETH Zurich, Clasiusstrasse 50, 8092 Zurich, Switzerland
\and 
Department of Transportation Networks, University of Zilina, Univerzitna 8215/5, 01026 Zilina, Slovakia 
\and
Departament de F\'{\i}sica Fonamental, Universitat de Barcelona, 08028 Barcelona, Spain
}
\date{Receiveed: date / Revised version: date}
%
\abstract{
We study synchronization dynamics in networks of coupled oscillators with bimodal distribution of natural frequencies. This setup can be interpreted as a simple model of frequency synchronization dynamics among generators and loads working in a power network. We derive the minimum coupling strength required to ensure global frequency synchronization. This threshold value can be efficiently found by solving a binary optimization problem, even for large networks. In order to validate our procedure, we compare its results with numerical simulations on a realistic network describing the European interconnected high-voltage electricity system,  finding a very good agreement. Our synchronization threshold can be used to test the stability of frequency synchronization to link removals. As the threshold value changes only in very few cases when aplied to the European realistic network, we conclude that network is resilient in this regard. Since the threshold calculation depends on the local connectivity, it can also be used to identify critical network partitions acting as synchronization bottlenecks. In our stability experiments we observe that when a link removal triggers a change in the critical partition, its limits tend to converge to national borders. This phenomenon, which can have important consequences to synchronization dynamics in case of cascading failure, signals the influence of the uncomplete topological integration of national power grids at the European scale.
} 
\maketitle
\section{Introduction}
\label{intro}
The electrical power grid is an example of complex system relying on the proper interaction between a great number of elements~\cite{ps_as_complexsystems}. Recently it attained  considerable attention of the complex systems community due to many features as, for instance, its non-trivial topological properties~\cite{Sole2008,buzna2009a,Rosato2005}, the presence of cascading effects~\cite{motter2002,Buzna2008,Crucitti2004,compara_models,Qioung2005}, self-organized criticality as a possible explanation for the frequency of blackouts~\cite{ieee_SOC}, its interaction with  other network systems~\cite{interdependent_nets,Bloomfield2010}, and also synchronization phenomena~\cite{filatrella2008,buzna2009}. Besides, current organizational trends are posing new research challenges related to the control of electric power grids. Two main prevaliling tendencies are the progressive integration of national networks into a European-wide one \cite{Zhou2005}, and the shift from centralized energy production towards more decentralized smart grids \cite{Schlaepfer2010}.

Power systems are formed by a large number of generators interconnected in a complex pattern to supply energy to final consumers. Modeling the complete set of variables that characterizes the whole system represents a tantalizing effort. Here we follow the direction to reduce the complexity of individual elements in favor of the complexity of the interaction pattern.

Our point tries to be a compromise of two opposite directions of research. On the one hand a more technical perspective (electrical engineering) relies on the complicated (not complex) arrangement of primary, secondary, and tertiary circuits to analyze the frequency stability of the generators~\cite{ps_stability,bergen1981}. On the other hand, a more statistical view (network science) analyzes the global behavior mainly from the topological static features  of the network; hence, the role of the different elements that form the networks (nodes for generators and loads, links for distribution lines) is uniquely associated with topological measures as, for instance, different types of centrality (local or global)~\cite{latora2001,albert2004}. As discussed in~\cite{integra_models} these two lines of research can usefully inform each other in order to bring new progress into an integrated study of very large systems formed by units with internal complex dynamics.

This paper can be also seen as an extension of the work presented in ~\cite{buzna2009}, since here we are providing a general formalization of the synchronization threshold where the previous relation was a special case. Moreover, in the present work we apply this generalized approach to study the influence of topological features on the stability of frequency synchronization in a realistic system.

In section~\ref{sec:equations} we start the paper  by reviewing the dynamical description of power systems as networks of Kuramoto oscillators~\cite{kuramoto1984}. In section~\ref{sec:stability} we derive a new formula describing the critical coupling strength and test its validity using realistic network topologies. In section~\ref{sec:simus} we analyse the sensitivity of frequency synchronization to link removals, and our findings and conclusions are summarized in section~\ref{conclusions}. 
\section{From swing and flow equations to oscillatory dynamics}
\label{sec:equations}

Our aim is precisely to incorporate simple dynamical behavior of generators, i.e. not to deal with a complete dynamical description involving electromagnetic fields and flow equations, but considering only the basic mechanism that ensures one of the crucial dynamical properties of the power distribution system as a whole, namely its synchronization. Synchronization  is understood as the ability of an (unsupervised) system to keep a global state where generators and loads (with intrinsic different frequencies) run with the same effective frequency (that of the distribution system, 50/60 Hz). To this end, generators and loads are described as phase oscillators in terms of "swing" equations. As it has been shown recently, this equation can be derived from a proper energy balance in the generator~\cite{filatrella2008}. The mechanical/thermal power brought to the generator gives rise to different contributions
\be
P_{source}=P_{dissipated}+P_{accumulated}+P_{transmitted}
\ee
where 
\be
P_{dissipated}=\gamma \dot{\theta}^2
\label{diss}
\ee
corresponds to the rate at which energy is dissipated due to the rotation of the mechanical rotors (turbines) and $\gamma$ is a damping coefficient,
\be
\label{acc}
P_{accumulated}=I
\frac{d\dot{\theta}}{dt}\dot{\theta}=I\dot{\theta}\ddot{\theta}
\ee
corresponds to the rate at which kinetic energy is accumulated, with $I$ being the inertia moment of the mechanical rotor. In Eqs.~(\ref{diss})-(\ref{acc}) $\theta$ is the angle of the mechanical rotor. Finally, the energy is transmitted to the distribution system through the lines. The transmitted power is proportional to the sinus of the phase difference between the voltages of the elements at the two end-points of the line. Here it lies one of the crucial points of the mechanical-electrical approach, i.e. the assumption that the voltage angles and the rotor angles are the same. Thus, we can write the power transmitted from element $i$ to $j$
\be
P_{ij}=-P_{ij}^{MAX} \sin (\theta_j -\theta_i),
\label{trans}
\ee 
with $P_{ij}^{MAX}$ being the maximum power transferred between generators $i$ and $j$~\cite{dorfler2011}.
Hence, we can write for each element
\be
P_i=\gamma_i \dot{\theta}_i^2 + I \frac{d\dot{\theta_i}}{dt}\dot{\theta_i} -
\sum_j 
P_{ij}^{MAX} \sin (\theta_j -\theta_i)
\label{powers}
\ee
assuming that $P_{ij}^{MAX}=0$ if elements $i$ and $j$ are not physically connected. This equation can describe generators as well as loads with the only difference that for loads $P_i<0$. The frequencies of the different elements cannot be very far from the standard frequency of the distribution system ($\Omega$=50/60 Hz), then we write
\be
\theta_i = \Omega t + \varphi_i
\ee
assuming $\dot{\varphi}_i\ll \Omega$. Inserting this expression in~(\ref{powers}) and keeping linear terms in  $\dot{\varphi}_i$ we get
\be
\dot{\varphi}_i= \left[ \frac{P_i}{2\gamma_i \Omega}-\frac{\Omega}{2} \right] -\frac{I_i}{2\gamma_i } \ddot{\varphi}_i + \frac{1}{2\gamma_i \Omega}\sum_j P_{ij}^{MAX}\sin (\varphi_j -\varphi_i)
\label{eq_filla}
\ee
which, for simplicity, can be recast as
\be
\dot{\varphi}_i=\omega_i -\alpha_i \ddot{\varphi}_i +\sum_j w_{ij} \sin
(\varphi_j -\varphi_i),
\label{eq_KM-like}
\ee
having introduced $\omega_i$ as the natural frequency of unit $i$ and $w_{ij}$ the coupling term.
The simple form of  Eq.~(\ref{eq_KM-like}) recalls the swing equation and this is the reason for its name in the technical literature, but the forcing term is what links the evolution of the phases of the element to the distribution system. In terms of dynamical systems and complex networks the meaning of Eq.~(\ref{eq_KM-like}) is obvious. It represents the evolution of phase oscillators with an additional inertia term~\cite{PRE_Acebron,RevModPhys_Acebron} embedded in a complex network \cite{adkmz08} whose elements interact in a weighted way through the sinus of the phase differences between the end points of a physical distribution line.

\textbf{The Kuramoto model with inertia has been analyzed previously in the literature and its main effect is on the type of transition that changes from second to first order when the coupling is increased. Although in \cite{PRE_Acebron} it is reported that inertia modifies the phase diagram of the system, one has to bear in mind that in that case the authors consider also the effects of noise in the system. For noiseless systems, as those considered in \cite{PhysD_Tanaka,PRL_Tanaka}, it is shown that there are two critical values of the coupling strength and a hysteresis behavior within this range of values. The lower value, below which the incoherent state prevails, is not modified with respect to the original Kuramoto model; but, the upper value, above which the synchronized state is the only possible solution, changes with the inertia contribution. In the intermediate region, there is a coexistence of the synchronized and incoherent states. The hysteresis behavior shows that the stability of the synchronized state is broken at the lower value of the coupling strength, and this is the important effect we will look at in this work. Under usual operational conditions the system is frequency (not phase) synchronized and we want to analyze how any disturbance (removal of some links) can affect this state, this means that only the lower value of the coupling strength matters, since it is the value below which the synchronized state is not achievable. Additionally, it has been shown \cite{Choi201132,pricomm} that the inertia term modifies the time scale for the achievement of the synchronized state but not the existence of the state. For all these reason we will remove the inertia term, which only speeds up convergence of the numerical simulations but it does not affect the conditions under which the system can achieve synchronization.}
\section{Stability of frequency synchronization}
\label{sec:stability}
What makes a systems of oscillators interesting is its dynamical behavior and, from a collective point of view,  the most relevant feature is its ability to synchronize. As we will show in this section, for a population of non-identical oscillators, perfect phase synchronization is not possible, whereas frequency synchronization requires a minimum coupling strength.
\subsection{Estimation of the minimum coupling required}
\label{subsec:estimation}

As it was shown in~\cite{buzna2009}, when we neglect the inertia term from Eq.~(\ref{eq_KM-like}), we can relate the synchronization threshold to the network topology. We consider the networked population of oscillating units, organized in a graph ${\cal G}$ composed from a set of nodes $N$ and a set of links $L$. Each unit (node) $i \in N$ is characterized by a natural frequency $\omega_i$  and a phase angle $\varphi_i$. The dynamics of these  units is then governed by
\be
\dot{\varphi_i} = \omega_i + \sigma \sum_{j} a_{ij} \sin(\varphi_j - \varphi_i),
\label{eq_KM}
\ee
where $a_{ij}$ is an element of the adjacency matrix which takes value $1$ when nodes $i$ and $j$ are connected by a link and value $0$ otherwise. Note that compared to Eq.~(\ref{eq_KM-like}), we have neglected the inertia term, we assume all links to be identical, and their coupling strength is characterized by a parameter $\sigma$. Motivated by the existence of generators and loads in power systems, we split the nodes into two populations of identical elements, where the members of the first population $N_+ \subseteq N$ behave like power producing units, and thus $\omega_i = \omega_{+} > 0$ for $i \in N_+$. And vice versa, the members of the second population $N_- \subseteq N$, where $N_- \cup N_+ = N$ and $N_- \cap N_+ = \emptyset$, behave as power consuming units and $\omega_i = \omega_{-} < 0$ for $i \in N_-$. Since the number of generators and the number of loads are not necessarily equal, we normalize $\omega_i$ values in the following way $ \omega_{+} = 1/|N_+|$ and $\omega_{-} = -1/|N_-|$. The choice of the positive sign for the natural frequency of the generators is not arbitrary; note that in eq. (\ref{eq_filla}) this is the case whereas for loads this term is clearly negative.

In general, for a population of Kuramoto oscillators there is a balance between the two terms in eq. (\ref{eq_KM}). On the one hand, if there is no coupling all units follow their natural frequencies $\omega_i$ and there is no frequency synchronization. On the other hand, when the coupling is very large the effective frequencies $\dot{\varphi}_i$ tend to zero. Then, there will be a critical value of the coupling strength $\sigma$ above which the system synchronizes in the sense that the effective frequencies become equal. Note that this frequency synchronization does not imply phase synchronization, which is only possible, for a population of non-identical oscillators, when the coupling strength goes to infinite.

Following the derivations presented in~\cite{buzna2009}, we can write for any unit in the synchronized state
\be
0 = \omega_i + \sigma \sum_{i,j} a_{ij} \sin(\varphi_j - \varphi_i).
\label{eq_EQ2}
\ee
Here it is easy to see that a necessary, but not sufficient, condition for the natural frequency of unit $i$ being 0 is that 
\be
\sigma>\sigma^c_i=\frac{|\omega_i |}{k_i}
\label{crit_node}
\ee
where $k_i$ is the degree of the node $i$. From this expression it is, in principle, possible to find a global bound for the coupling strength. This bound can be, however, far from the real value, since there are some geometric constraints \cite{luce_pre} (phase differences that cannot be maximized simultaneously) involving more than one unit. Actually, we can write as many equations as partitions of the network into two non-overlapping clusters, just by summing the equations for the nodes and using the fact that the interaction is an odd function. Identifying the network partition as a set of nodes $S$, we can write for the sum of the Eqs. of the nodes in $S$
\be
0 =  \sum_{i \in S}\omega_i+ \sigma \sum_{i \in S, j \notin S} a_{ij} \sin(\varphi_j - \varphi_i).
\label{eq_EQ1}
\ee
from which we can find the critical values of the coupling for the partition $S$
\be
\sigma>\sigma^c_S=\frac{\left|\sum_{i \in S}\omega_i\right|}{\sum_{i \in S, j \notin S} a_{ij} }
\ee
which generalizes (\ref{crit_node}), containing it as particular cases of clusters formed by individual nodes. Note that in~\cite{buzna2009} we restricted ourselves to clusters of the same type of node, which gives good estimations for homogeneous  distributions, but not for a more even distribution as is the current case of interest. The goal is to obtain the global minimum that ensures the existence of a synchronized state
\be
\label{eq:sigma}
\sigma>\max_{S} \frac{\left|\sum_{i \in S}\omega_i\right|}{\sum_{i \in S, j \notin S} a_{ij} }
\ee
\subsection{Testing numerically the critical coupling}
\label{subsec:numeric}
To test our hypothesis we use the approximative data describing the topology of the European power transmission network~\cite{Qioung2005} composed from sixteen national subnetworks. The network topology was derived from maps published by UCTE~\cite{ucte}. Functional parameters, as for example position of power producing and power consuming units, their capacities and capacities of power lines, were estimated in order to fit the publicly known volumes of cross border flows as precisely as possible.
 
\begin{figure}
\includegraphics[width=0.48\textwidth]{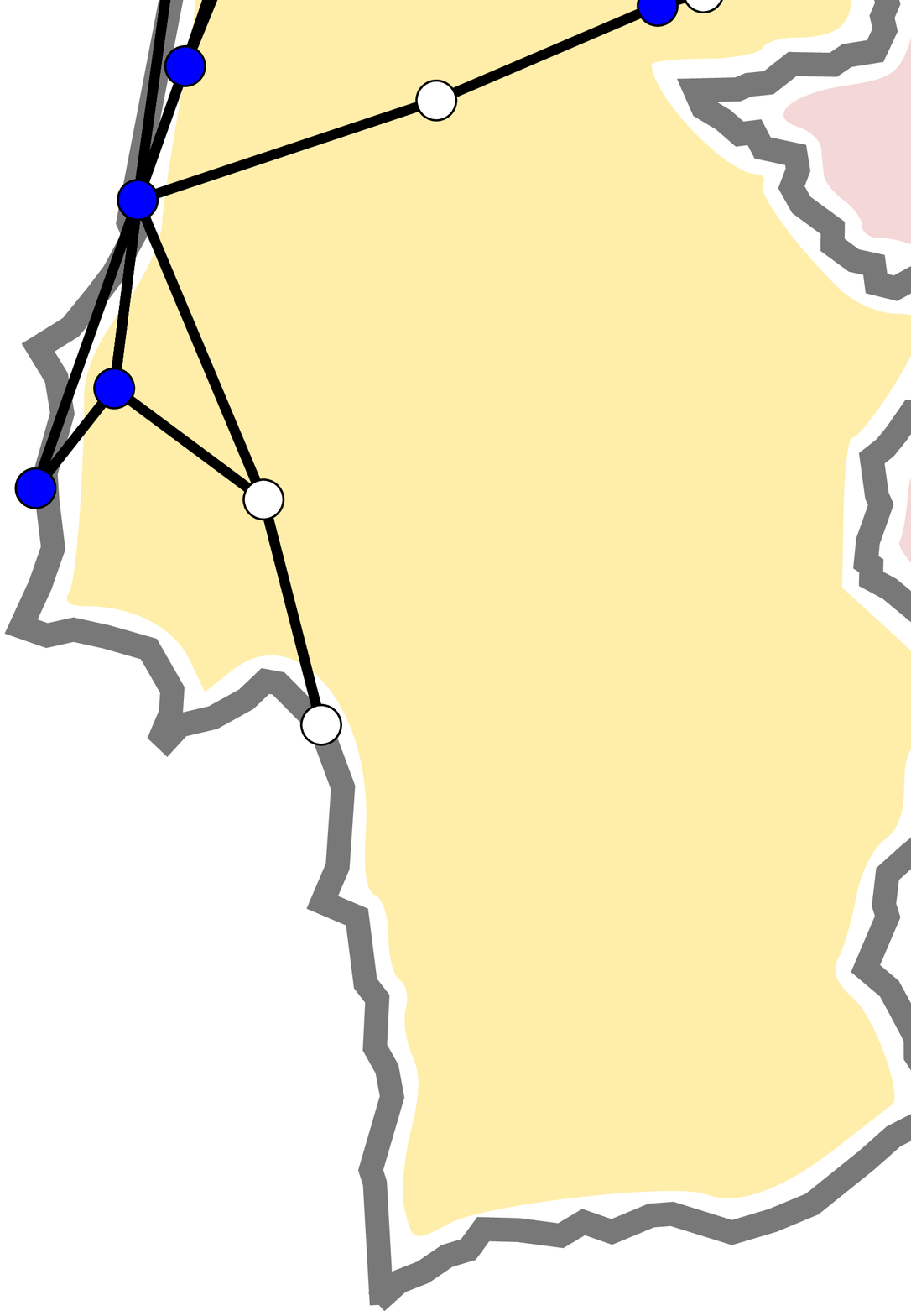}
\caption{Topology of the approximate model of European interconnected power system~\cite{Qioung2005}. White circles represent positions of active (power injecting) stations and blue dots stand for passive (power consuming) stations.}
\label{fig:data}
\end{figure}

The network is depicted in Fig.~\ref{fig:data}, where white circles represent positions of active (power injecting) units and blue dots stand for passive (power consuming) units (can be seen online when the figure is sufficiently zoomed). The first two columns in Table~\ref{tab:tau} summarize the number of vertices and edges for the European network as a whole, and for each one of the national subnetworks (i.e. those parts of the network contained within each country boundaries).

According to the previous paragraphs, we have applied Eq.~(\ref{eq:sigma}) to each one of the networks in order to determine their critical coupling value $\sigma_c$ (i.e. the minimum one needed to have frequency synchronization)\footnote{Absolute values of $\sigma_c$ are fully dependent on $\omega_{+}$ and $\omega_{-}$. The values of $\omega_{+}$ and $\omega_{-}$ were chosen in a way that they sum up to 1 and -1, respectively. As the number of nodes in national networks differs it makes the comparison between countries difficult.}. Specifically, we have applied a semi-analytical approach to determine the partition within the network maximizing the expression in Eq.~(\ref{eq:sigma}). This procedure provides lower and upper bounds for $\sigma_c$, $\sigma_A^l$ and $\sigma_A^u$, respectively. The third and forth columns in Table~\ref{tab:tau} present $\sigma_A^l$ and $\sigma_A^u$ for each one of the considered networks. More details on the methodology can be found in the Appendix.
 
To verify the goodness of these values as approximations to the critical coupling, we have to check whether they are the minimum value of $\sigma$ allowing full synchronization of the system. We have done so numerically by simulating frequency synchronization dynamics. More concretely, we have run synchronization dynamics described in Eq.~(\ref{eq_KM}) and measured frequency dispersion ($r$), which is the order parameter proposed in~\cite{buzna2009} \textbf{to measure of the effective frequency dispersion}:\\

\begin{equation}
\label{eq:calculation_sigma}
r = \sqrt{\frac{1}{N}\sum_{i \in N}[\dot{\varphi}_i-\langle\omega\rangle]^2},
\end{equation}

\begin{figure}
\includegraphics[width=0.48\textwidth]{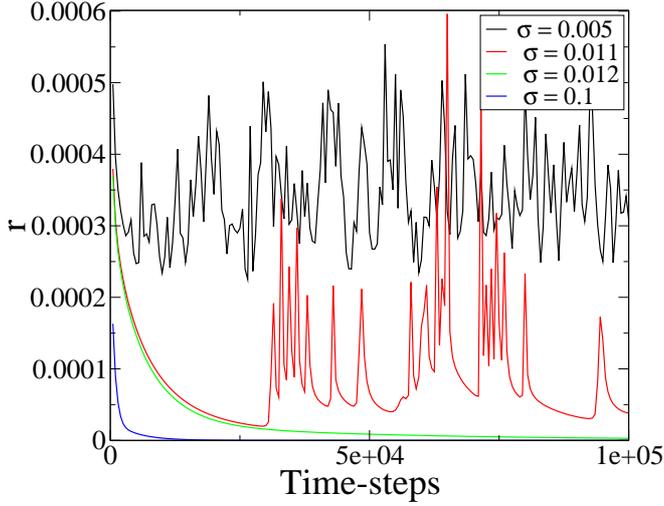}
\caption{Temporal evolution of frequency synchronization in the European network for different coupling values. Frequency dispersion $r$ (the order parameter introduced in~\cite{buzna2009}) is used to show the effect of coupling $\sigma$ below (0.005 and 0.011) and above (0.012 and 0.1) the critical value. Notice that the plot corresponding to $\sigma=0.011$ presents a characteristic behavior of a system very close to criticality. First it relaxes towards synchronization but, at a certain point, it experiences sharp fluctuations that are stronger than those in the case $\sigma=0.005$. Simulations start with all phases set to 0 so, as the dynamics are purely deterministic, only one realization per coupling value was needed.}
\label{fig:numerical_european}
\end{figure}

Fig.~\ref{fig:numerical_european} presents time evolution of the order parameter $r$ for different values of the coupling $\sigma$ for the European network. For values far enough from the critical one, the system either fluctuates steadily around a certain value (see $\sigma=0.005$ in the figure) or relaxes towards full frequency synchronization ($\sigma=0.1$ case). As we get closer to the critical value, we observe an initial tendency toward synchronization that is sharply broken by very strong fluctuations ($\sigma=0.011$).\\

For each network we have executed an iterative procedure running the dynamics with several coupling values and checking whether $r$ was relaxing towards 0. In each case, the minimum coupling satisfying this condition \textbf{, $\sigma_N$,} has been determined with a precision of four decimal places.

\begin{figure}
\includegraphics[width=0.48\textwidth]{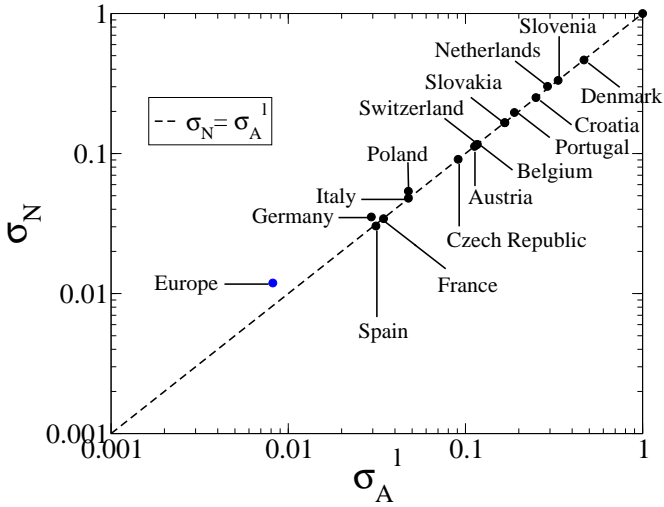}
\caption{Comparison of $\sigma^l_A$ values determined by means of Eq.~(\ref{eq:sigma}) and  $\sigma_N$ obtained numerically using the iterative approximation method. As a guide to the eye, we have added a dashed line corresponding to $\sigma_N=\sigma_A^l$. Clearly, there is a remarkable agreement. This result supports our claim that Eq.~(\ref{eq:sigma}) is a good estimation of the minimum coupling needed to assure frequency synchronization.}
\label{fig:data1}
\end{figure}

\textbf{Finally, ${\sigma^l}_A$ values in Table~\ref{tab:tau} (obtained analytically applying Eq.~(\ref{eq:sigma})) can be checked by comparing them with these $\sigma_N$'s  obtained numerically. Fig.~\ref{fig:data1} presents all values together to allow for a visual comparison.} Generally speaking, we observe a remarkable agreement between the two sets of results. Even for the three cases where the difference among analytical and numerical approximations is visible (i.e. Europe as a whole, Germany and Poland), the obtained values are of the same magnitude.

This outcome supports our claim that Eq.~(\ref{eq:sigma}) is a good estimation of the critical coupling $\sigma_c$, and opens the door to simple studies on frequency synchronization in networks. In particular, Eq.~(\ref{eq:sigma}) can be applied to the study of frequency synchronization's stability against perturbations.
\begin{table}[h]
\begin{center}
\begin{tabular}{|r|r|r|r|r|}
\hline
Country Name & N & L &  $\sigma_A^l$ & $\sigma_A^u$  \\
\hline
Europe      &	1254 &	1943  & 0.00818157 & 0.00818253\\ 
Austria     &	36 &	429 & 0.112499 & 0.1125\\  
Belgium     &	22 &	21  & 0.116666 & 0.116667 \\ 
Croatia     &	17 &	20  & 0.25 & 0.250001\\
Czech Republic &	34 &	52  & 0.090909 & 0.09091\\
Denmark     &	8 &	8  & 0.466666 & 0.466667\\
France      &	318 &	519 & 0.034482 & 0.034483\\ 
Germany     &	229 &	313  & 0.0294113 & 0.0294123\\
Hungary     &	27 &	36  & 0.166666 & 0.166667\\
Italy       &	139 &	204  & 0.0476074 & 0.0476084\\ 
Luxemburg   &	3 &	2 & 0.999999 & 1\\
Netherlands &	22 &	24  & 0.290598 & 0.290599\\
Poland      &	99 &	140  & 0.043478 & 0.043479\\
Portugal    &	24 &	44  & 0.188889 & 0.18889\\
Slovakia    &	25 &	30  & 0.166666 & 0.166667\\
Slovenia    &	8 &	8  & 0.333333 & 0.333334\\  
Spain       &	193 &	316  & 0.03125 & 0.031251\\
Switzerland &	47 &	76  & 0.115384 & 0.115385\\
\hline
\end{tabular}
\caption{Network size and the values of $\sigma^l_A$ and $\sigma^u_A$ for the whole European high-voltage electrical network and the national networks. Values $\sigma_A^l$ and $\sigma_A^u$ have been calculated according to Eq.~(\ref{eq:sigma}) following the procedure described in the Appendix.}
\label{tab:tau}
\end{center}
\end{table}

\section{Resilience of power systems in terms of frequency synchronization stability}
\label{sec:simus}
 
There is a bulk of literature analyzing the robustness of power systems. Since power systems can be represented as networks, most of these approaches deal with the robustness of the system to the removal of links or vertices. This kind of robustness analysis has been addressed in many different ways. Initial purely topological approaches~\cite{Albert:pg-topo} have shown to be limited ~\cite{compara_models}, and recent ones combine structural changes with different processes such as flow dynamics~\cite{Buzna2008,Kinney:pw-cascading}.

Following up from this literature, we address the robustness of power systems to link removal from the viewpoint of the stability of frequency synchronization. A straightforward way to make such an analysis is to calculate to what extent the removal of a link from the network modifies the critical value of the coupling  $\sigma_c$ for the whole system.
 
More concretely, the idea is to test whether removing certain links from the network can increase the minimum coupling value required to ensure synchronization (i.e. making it less stable) or the other way around, decrease it (subsequently improving its robustness).

\subsection{Effect of link removal on synchronization stability}
\label{subsec:static_link_removal}

\begin{figure}
\includegraphics[width=0.48\textwidth]{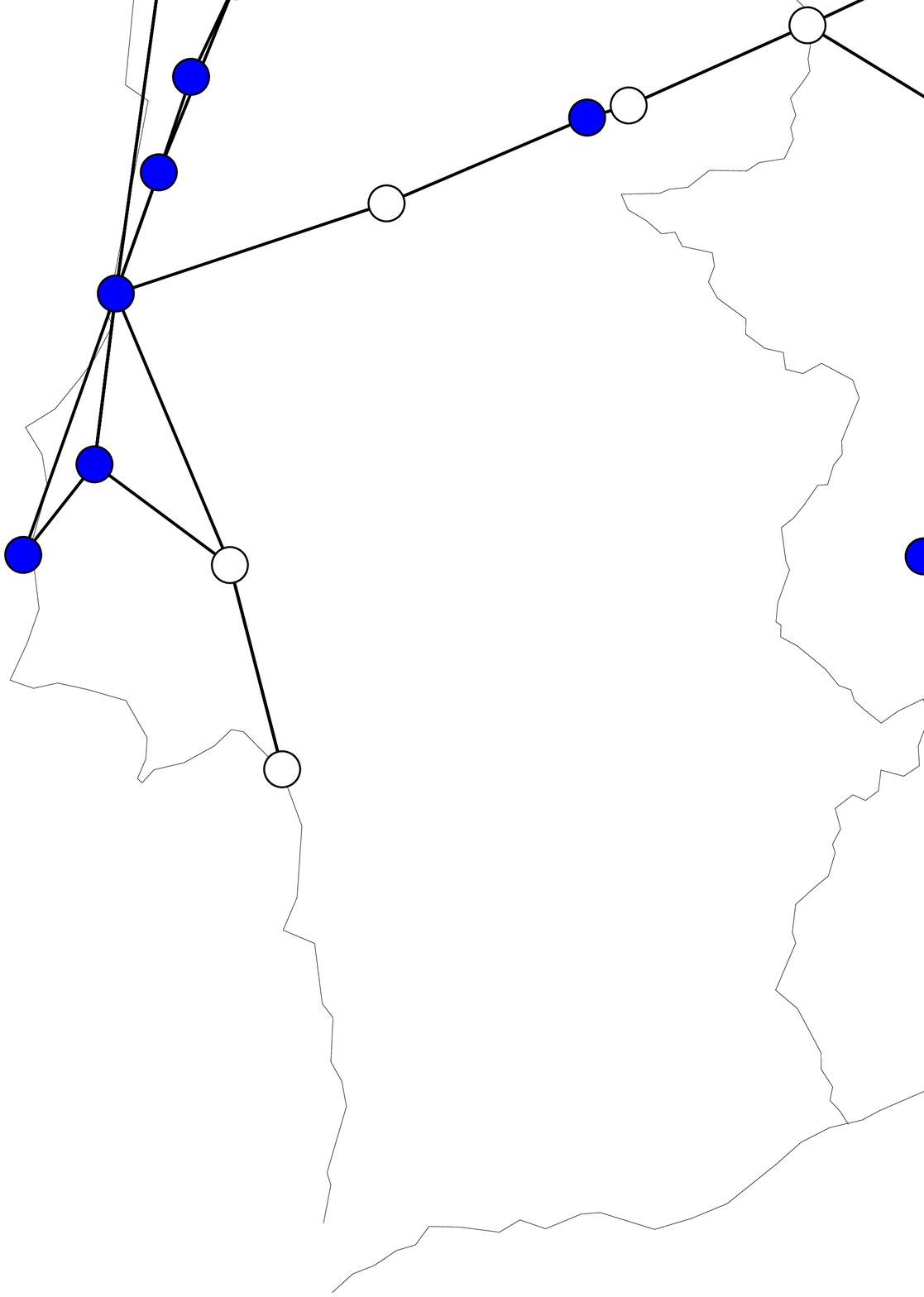}
\caption{Critical partition of the European network. The shadowed area corresponds to the partition defining the critical value of coupling ($\sigma_A^l$) when the European network is complete, and the links marked in green connect it to the rest of the network. Notice that, given the conditions stated in section 2, the complementary partition (i.e. the whole network except the shadowed area) presents the same critical coupling, so any of the two could be considered the critical one.}
\label{fig:data4}
\end{figure}

Our starting point is the complete European network. Fig.~\ref{fig:data4} shows the location of the partition maximizing Eq.~(\ref{eq:sigma}) and, therefore, determining $\sigma_c$. Then, for each one of the links in the European network $e$, we take it out from the grid, calculate the new value of $\sigma_A^l(e)$, and restore it to its position.

\begin{figure}
\includegraphics[width=0.48\textwidth, angle = 0]{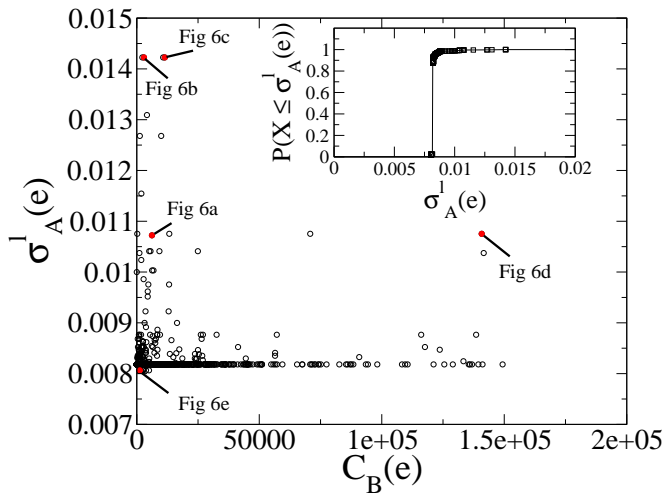}
\caption{Impact over global frequency synchronizity of removing each edge of the European Network. New critical coupling values ${\sigma^l}_A$ were obtained from Eq.~(\ref{eq:sigma}) as described in the Appendix, and edges are ordered according to their Edge Betweenness $C_B(e)$. Inset: Cumulative probability distribution of~${\sigma^l}_A$ values. The network is robust to removal of all but very few edges, whose actual impact when taken out is not correlated with their centrality.}
\label{fig:data3}
\end{figure}

As shown in Fig.~\ref{fig:data3}, only few links in the whole European network have an impact on the global $\sigma_c$ value when removed. Among them, a minority decrease~$\sigma_A^l$ (i.e. make overall frequency synchronization easier if erased), while the other ones exercise the opposite influence, since their removal increases~$\sigma_A^l$ value.
 
We can also observe that such effects of link deletion are not related to link betweenness~\cite{brandes2005}. This result is especially relevant, since purely topological measures (such as link betweenness or degree centrality) have been traditionally used to assess resilience in power grids and other networked systems. On the contrary, here a strictly topological approach is no longer valid and we need to include dynamical aspects in our analysis. Accordingly, understanding the effects shown in Fig.~\ref{fig:data3} requires interpreting how a link removal affects synchronization dynamics in Eq.~(\ref{eq_KM}) and, as a consequence, modifies the global critical coupling.

\begin{figure}
\includegraphics[width=0.48\textwidth]{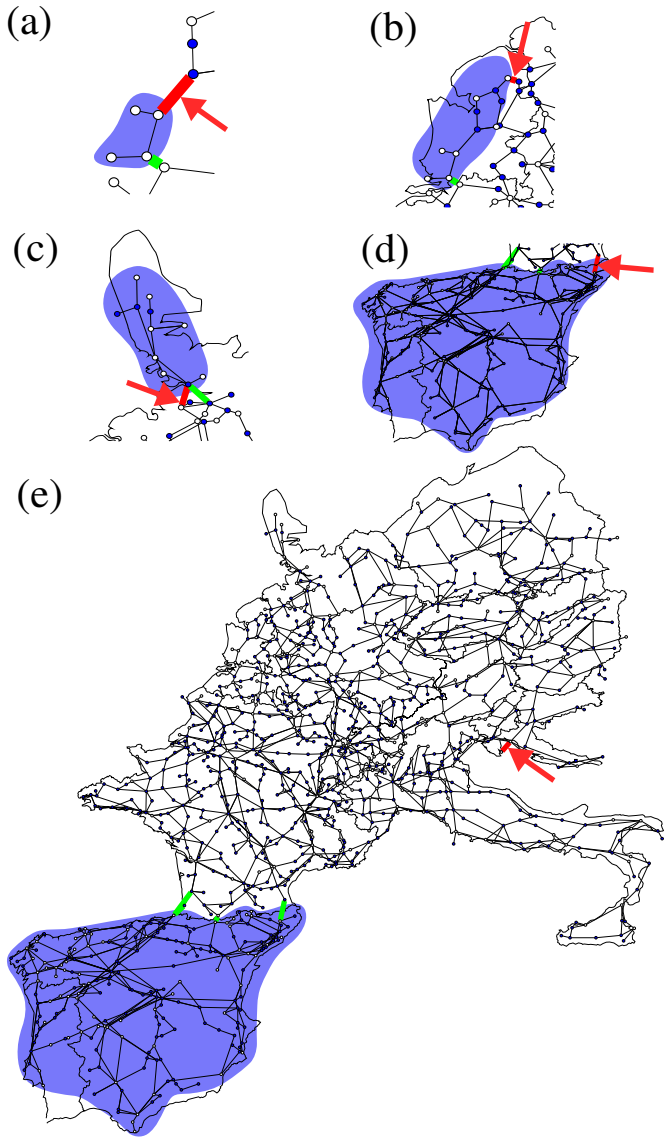}
\caption{Examples of link removals modifying the critical value of coupling $\sigma_A^l$ of the European network. In cases a) - d), the link removal increases $\sigma_A^l$, while it is reduced in case e). For all cases, removed links are marked in red and indicated with an arrow, the new critical partition is marked as a shadowed area and the links connecting it to the rest of the network are marked in green. More details are provided in the main text.}
\label{fig:data5}
\end{figure}

Fig.~\ref{fig:data5} provides such a detailed view of the local structure for five of the links which deletion changes the global critical coupling. Removed links are marked in red, the shadowed areas correspond to the partition becoming critical after the removal, and green ties are their remaining connections with the rest of the network.

Figs.~\ref{fig:data5}(a)-(d) show cases where the deletion of the link marked in red makes the critical coupling $\sigma_c$ to increase. It is quite intuitive to see that the removed links were partition's 'boundary' links (i.e. pointing to outside of the partitions) and, therefore, their removal reduces the connections of the partitions to the rest of the network (e.g. in case~\ref{fig:data5}(a), for instance, from two to just the one marked in green). In accordance with Eq.~(\ref{eq:sigma}), this reduction leads to an increase of the partition's critical coupling, which eventually becomes the new global $\sigma_c$.

Fig.~\ref{fig:data5}(e) presents an example of the opposite case (i.e. $\sigma_c$ diminishing). Here the erased link was an inner one, connecting a leaf (i.e. a node of degree one) to the rest of the partition. Therefore its deletion reduces the size of the partition and, again according to Eq.~(\ref{eq:sigma}), makes the critical value of the coupling for the partition smaller. This reduction then let other partitions in the network with a critical coupling lower than the previous global $\sigma_c$ become the new critical partition. In this particular case, the new critical partition corresponds to the Iberian Peninsula. Notice that in contrast with case~\ref{fig:data5}(d), which presents the same resulting critical partition, in this case the number of remaining boundary links are 3, so the critical coupling of the partition is lower.

\section{Conclusions}
\label{conclusions}
We  studied how synchronization dynamics of networked oscillators depend on their topological configurations. When neglecting control mechanisms present in real power systems and considering only some aspects of the swing and flow equations, oscillatory dynamics can be captured by the Kuramoto model assuming a bipolar distribution of natural frequencies. For this setup we are here giving accurate analytical estimation of the critical coupling strength. It leads to the solving of a combinatorial optimization problem on the graph. This can be efficiently done using methods of binary linear programming even for very large networks with an arbitrary precision. We validated our results on realistic network topologies and by comparing them with numerical simulations we found very good agreement. 

Furthermore, we studied the robustness of the synchronization process with respect to removals of single links. When links are removed, only a small percentage of them have an impact on the synchronization threshold. Thus network is in this respect very robust. While some links are increasing the synchronization threshold the others are decreasing it. The concrete effect depends on whether they are decreasing the size of the critical cluster or affecting its connectivity with the rest of the network. We also tested whether the positions of links influencing the synchronization threshold is correlated with their global topological properties as is sometimes assumed when assessing the resilience of real-world network topologies. Here we find no correlation. This can be explained when interpreting the Eq.(\ref{eq:sigma}) which  estimates the synchronization threshold. For a given subset of nodes the threshold value is higher the higher is the sum of natural frequencies for given subset of nodes and the less tightly are these nodes connected with the rest of the network. Thus the synchronization threshold depends on the local properties of the critical network partition. 

When analyzing the geographical positioning of critical partitions, we often found them being identical to borders  between European countries. For example, for the original network without removed links the critical clusters are connected along the western French border (see Fig.~\ref{fig:data4}). When links are removed, in some cases critical clusters are connected along the Spanish-French border (see Figs.~\ref{fig:data5}d and~\ref{fig:data5}e). Another example is the critical partition corresponding to the Dutch network (see Fig. \ref{fig:data5}b). These observations indicate the tendency of national networks to be more tightly connected internally than across borders, which is a signature of an on-going (incomplete) European-wide integration process. This fact can have positive consequences from a practical point of view, as it leads to a natural tendency in the network to split along these areas in critical situations. The major disruption in the synchronization  of the European network in November 2006, which resulted in a temporal division of the system into three regions mainly following existing or former national boundaries, is an illustrative example of this dependence \cite{UCTE_report}.

This paper is contributing to a better understanding of the interplay between the network topology determining the spatial positioning of network elements and frequency synchronization dynamics. Future steps could include the study of the cascading behaviours, whereas the Kuramoto equations allow to define link flows, or  investigation of the validity of our results when considering a more realistic model, e.g. by introducing heterogeneous links or various types of generators and loads. \textbf{Finally, this article was focused on the existence and stability of frequency synchronization, where the inertia has a limited effect. Possible extensions of our work could address, for instance, the non-trivial influence of the inertia term on transient states pointed out in section 2.}

\section*{Acknowledgments}
L.B.  gratefully acknowledges partial financial support by the Grant agency of the Ministry of Education of the Slovak Republic and the Slovak Academy of Sciences (project VEGA 1/0296/12).
A.D-G. has been supported by Ministerio de Educación y Ciencia (PR2008-0114), by the Spanish DGICyT (FIS-2006-13321 and FIS2009-13730), and by the Generalitat de Catalunya 2009SGR00838.

\section*{Appendix: Methodology}
\label{appendix}
The problem of identifying the subset of nodes maximizing Eq.~(\ref{eq:sigma}) can be efficiently solved by using  binary linear programming~\cite{Nemhauser1988}. First, we formulate it as a combinatorial optimization problem: \textit{Let us have a graph $\mathcal{G}(N,L)$ where $N$ is a set of nodes and $L$ is a set of links. Each node has a real value $\omega_i$ associated to it. The problem is to find a nonempty subset of nodes $S \subset N$  maximizing the expression $\frac{\sum_{i \in S} \omega_i}{\sum_{i \in S, j \notin S} a_{ij}}$.}

Thus, we need to decide which nodes are, and which nodes are not, included in the subset $S$. Such decision can be modeled by a set of binary variables $y_i \in \{0, 1\}$ for $i \in N$. Variable $y_i = 1$ iff $i \in S$ and otherwise $y_i = 0$. In the denominator of Eq.~(\ref{eq:sigma}), we consider the number of links connecting nodes belonging to  the set $S$ with nodes which do not belong to the set $S$. To count them, we introduce for each link connecting nodes $i$ and $j$ a binary variable $x_{i,j} \in \{0, 1\}$. Variable $x_{i,j}$ takes value $1$ iff either $i \in S$ and $j \in N-S$ or $j \in S$ and $i \in N-S$ and $x_{i,j} = 0$ otherwise. Then we can formally formulate this combinatorial optimization problem:

\begin{align}
\label{obj}
\text{Maximize }   & f =  \frac{\sum_{i \in N}{\omega_i y_i}}{\sum_{(i,j) \in L}{x_{i,j}}}   \\
\text{subject to} \nonumber \\
\label{con1}
 & \sum_{i \in N} y_i \geq 1 \\
\label{con2}
 &  x_{i,j} \geq y_j - y_i  & \text{ for } (i,j) \in L\\
\label{con3}
&  x_{i,j} \geq y_i - y_j  & \text{ for } (i,j) \in L\\
&y_i, \ x_{i,j} \in \{0,\ 1 \} &\text{ for } i \in N, (i,j) \in L
\end{align} 

Constraint~(\ref{con1}) ensures that the set $S$ is nonempty. The set of constraints~(\ref{con2}) and~(\ref{con3}) make sure that for each link connecting nodes $i$ and $j$ variable $x_{i,j} = 1$ if $y_i = 1$ and $y_j = 0$ or when $y_i = 0$ and $y_j = 1$. The objective function~(\ref{obj}) is forcing variables $x_{i,j}$ to take value of zero whenever it is possible. Therefore $x_{i,j}$ is zero if $y_i = 0$ and $y_j = 0$ or if $y_i = 1$ and $y_j = 1$. 

Note that we obtained an optimization problem with a non linear objective function~(\ref{obj}) that is optimized with respect to the linear set of constraints~(\ref{con1})-(\ref{con3}). When substituting the value of the objective function~(\ref{obj}) by the variable $\sigma$ we can rewrite the problem~(\ref{obj})-(\ref{con3}) as:

\begin{align}
\label{qobj2}
\text{Maximize }   & f =  \sigma\\
\text{subject to} \nonumber \\
\label{substitute}
 & \sigma \sum_{(i,j) \in L}{x_{i,j}} \geq \sum_{i \in N}{\omega_i y_i}\\
\label{qcon12}
 & \sum_{i \in N} y_i \geq 1 \\
\label{qcon22}
 &  x_{i,j} \geq y_j - y_i  & \text{ for } (i,j) \in L\\
\label{qcon32}
&  x_{i,j} \geq y_i - y_j  & \text{ for } (i,j) \in L\\
& \sigma \geq 0, y_i, \ x_{i,j} \in \{0,\ 1 \} &\text{ for } i \in N, (i,j) \in L
\end{align}

If we replace the variable $\sigma$ by an arbitrary constant $c$ we obtain a linear optimization problem which can be easily solved by traditional integer solvers as, for example, XPRESS-IVE~\cite{xpress}. The remaining problem has a feasible solution only if $c \leq \sigma_c$ and there is no solution if $c > \sigma_c$. Thus, by using a binary search on $c$ and repeatedly solving the optimization problem for different $c$ values we can find an arbitrarily tight lower $\sigma_A^l$ and upper $\sigma_A^u$ bounds for $\sigma_c$. Moreover, values of variables $y_i$ corresponding to the lower bound $\sigma_A^l$ (when feasible solution exists) define which nodes belong to the partition whose $\sigma = \sigma_A^l \leq \sigma_c < \sigma_A^u$.

\bibliographystyle{epj}
\bibliography{sync2}

\begin{thebibliography}{38}

\bibitem{ps_as_complexsystems}
I.~Dobson, B.~a~Carreras, V.E. Lynch, D.E. Newman, Chaos (Woodbury, N.Y.)
  \textbf{17}(2), 026103 (2007)

\bibitem{Sole2008}
R.V. Sol{\'e}, M.~Rosas-Casals, B.~Corominas-Murtra, S.~Valverde, Phys. Rev. E
  \textbf{77}, 026102 (2008)

\bibitem{buzna2009a}
L.~Buzna, L.~Issacharoff, D.~Helbing, IJCIS \textbf{5}(1/2), 72 (2009)

\bibitem{Rosato2005}
V.~Rosato, S.~Bologna, F.~Tiriticco, Electric Power Systems Research
  \textbf{77}(2), 99 (2007)

\bibitem{motter2002}
A.E. Motter, Y.C. Lai, Phys. Rev. E \textbf{66}, 065102 (2002)

\bibitem{Buzna2008}
I.~Simonsen, L.~Buzna, K.~Peters, S.~Bornholdt, D.~Helbing, Phys. Rev. Lett.
  \textbf{100}(21), 218701 (2008)

\bibitem{Crucitti2004}
P.~Crucitti, V.~Latora, M.~Marchiori, Phys. Rev. E \textbf{69}, 045104 (2004)

\bibitem{compara_models}
P.~Hines, E.~Cotilla-Sanchez, S.~Blumsack, Chaos (Woodbury, N.Y.)
  \textbf{20}(3), 033122 (2010)

\bibitem{Qioung2005}
Z.~{Qioung}, J.W. {Bialek}, IEEE Transactions on Power Systems \textbf{20}, 782
  (2005)

\bibitem{ieee_SOC}
B.~Carreras, D.~Newman, I.~Dobson, A.~Poole, Circuits and Systems I: Regular
  Papers, IEEE Transactions on \textbf{51}(9), 1733 (2004)

\bibitem{interdependent_nets}
S.V. Buldyrev, R.~Parshani, G.~Paul, H.E. Stanley, S.~Havlin, Nature
  \textbf{464}(7291), 1025 (2010)

\bibitem{Bloomfield2010}
R.~Bloomfield, L.~Buzna, P.~Popov, K.~Salako, D.~Wright, Lecture Notes in
  Computer Science \textbf{6027}, 201 (2010)

\bibitem{filatrella2008}
G.~Filatrella, A.~Nielsen, N.~Pedersen, EPJ B \textbf{61}(4), 485 (2008)

\bibitem{buzna2009}
L.~Buzna, S.~Lozano, A.~D{\'i}az-Guilera, Phys. Rev. E \textbf{80}(6), 066120
  (2009)

\bibitem{Zhou2005}
Q.~Zhou, J.~Bialek, IEEE Transactions on Power Systems \textbf{20}(3), 1663
  (2005)

\bibitem{Schlaepfer2010}
M.~Schl{\"a}pfer, K.~Trantopoulos, Phys. Rev. E \textbf{81}, 056106 (2010)

\bibitem{ps_stability}
P.~Kundur, J.~Paserba, IEEE Transactions on Power Systems \textbf{19}(2), 1387
  (2003)

\bibitem{bergen1981}
A.~Bergen, D.~Hill, Power Apparatus and Systems, IEEE Transactions on
  \textbf{PAS-100}(1), 25 (1981)

\bibitem{latora2001}
V.~Latora, M.~Marchiori, Phys. Rev. Lett. \textbf{87}(19), 198701 (2001)

\bibitem{albert2004}
R.~Albert, I.~Albert, G.L. Nakarado, Phys. Rev. E \textbf{69}(2), 025103 (2004)

\bibitem{integra_models}
D.~Hill, Circuits and Systems, 2006. ISCAS 2006. pp. 722--725 (2006)

\bibitem{kuramoto1984}
Y.~Kuramoto, \emph{Chemical Oscillations, Waves, and Turbulence}
  (Springer-Verlag, New York, 1984)

\bibitem{dorfler2011}
F.~Dorfler, F.~Bullo, SIAM Journal on Control and Optimization  (2011),
  submitted

\bibitem{PRE_Acebron}
J.A. Acebr\'on, L.L. Bonilla, R.~Spigler, Phys. Rev. E \textbf{62}, 3437 (2000)

\bibitem{RevModPhys_Acebron}
J.A. Acebr\'on, L.L. Bonilla, C.J. P\'erez~Vicente, F.~Ritort, R.~Spigler, Rev.
  Mod. Phys. \textbf{77}, 137 (2005)

\bibitem{adkmz08}
A.~Arenas, A.~D{\'i}az-Guilera, J.~Kurths, Y.~Moreno, C.~Zhou, Phys. Rep.
  \textbf{469}, 93 (2008)

\bibitem{PhysD_Tanaka}
S.O. Hisa-Aki~Tanaka, Allan J.~Lichtenberg, Physica D: Nonlinear Phenomena
  \textbf{100}(1), 279  (1997)

\bibitem{PRL_Tanaka}
H.A. Tanaka, A.J. Lichtenberg, S.~Oishi, Phys. Rev. Lett. \textbf{78}, 2104
  (1997)

\bibitem{Choi201132}
Y.P. Choi, S.Y. Ha, S.B. Yun, Physica D: Nonlinear Phenomena \textbf{240}(1),
  32  (2011)

\bibitem{pricomm}
Seung-Yeal Ha, private communication

\bibitem{luce_pre}
L.~Prignano, A.~D\'{\i}az-Guilera, Phys. Rev. E  (2012)

\bibitem{ucte}
\texttt{http://www.ucte.org}

\bibitem{Albert:pg-topo}
R.~Albert, I.~Albert, G.L. Nakarado, Phys. Rev. E \textbf{69}(2), 025103 (2004)

\bibitem{Kinney:pw-cascading}
R.~Kinney, P.~Crucitti, R.~Albert, V.~Latora, The European Physical Journal B -
  Condensed Matter and Complex Systems \textbf{46}, 101 (2005),
  10.1140/epjb/e2005-00237-9

\bibitem{brandes2005}
U.~Brandes, T.~Erlenbach, eds., \emph{Network Analysis: Methodological
  Foundations}, Lecture Notes in Computer Science (Springer-Verlag Berlin
  Heidelberg, 2005), ISBN 3-540-24979-6

\bibitem{UCTE_report}
UCTE, \emph{Final Report: System Disturbance on 4 November 2006} (UCTE, 2006)

\bibitem{Nemhauser1988}
A.W. {G.L. Nemhauser}, \emph{{Integer and Combinatorial Optimization}} (John
  Wiley \& Sohn, 1988)

\bibitem{xpress}
\texttt{http://www.fico.com}

\end{thebibliography}
\end{document}